\documentclass[preprint,11pt,5p]{elsarticle}

\usepackage{amssymb}
\usepackage{graphicx}
\usepackage{dcolumn}
\usepackage{bm}
\usepackage{slashed}
\usepackage{amsmath}
\usepackage{amsthm}

\usepackage{color}

\def\tr{\mbox{tr}\,}
\newcommand{\la}{\label}
\newcommand{\be}{\begin{equation}}
\newcommand{\ee}{\end{equation}}
\newcommand{\bea}{\begin{eqnarray}}
\newcommand{\eea}{\end{eqnarray}}
\newcommand{\p}{\partial}

\newcommand{\comment}[1]{}

\begin{document}

\begin{frontmatter}



\title{Entanglement Entropy in 2D Non-abelian Pure Gauge Theory }


\author{Andrey Gromov}
\address{Physics Department, Stony Brook University, Stony Brook, NY 11794-3840, USA}
\ead{andrey.gromov@stonybrook.edu}

\author{Raul A. Santos\corref{cor1}}
\address{C.N. Yang Institute for Theoretical Physics, Stony Brook University, Stony Brook, NY 11794-3840, USA}
\cortext[cor1]{Present Address: Department of Condensed Matter Physics, Weizmann Institute of Science, Rehovot, Israel; 
Department of Physics, Bar-Ilan University, Ramat Gan, 52900, Israel}
\ead{santos@insti.physics.sunysb.edu}

\begin{abstract}
We compute the Entanglement Entropy (EE) of a bipartition in 2D pure non-abelian $U(N)$ gauge theory. We obtain a general expression 
for EE on an arbitrary Riemann surface. We find that due to 
area-preserving diffeomorphism symmetry EE does not depend on the size of the subsystem, but only on the number of disjoint intervals defining the bipartition. 

 In the strong coupling limit on a torus we find that the scaling of the EE at  small temperature is given by $S(T) - S(0) = O\left(\frac{m_{gap}}{T}e^{-\frac{m_{gap}}{T}}\right)$, which is similar to the scaling for the matter fields recently derived in literature. In the large $N$ limit we compute all of the Renyi entropies and identify the Douglas-Kazakov phase transition.
\end{abstract}


\end{frontmatter}

\section{Introduction}
Entanglement has become a useful tool in the study of the properties of the states of matter \cite{amico2008a}. 
A particularly useful measure of entanglement in quantum systems is the Entanglement Entropy (EE). In a pure state, EE measures the entanglement present between a subsystem and its complement. When the subsystem is a subset of the configuration space (real space partition) EE can be computed 
using the replica trick \cite{Callan1994, calabrese2004a}. 

It was suggested on the basis of the AdS/CFT arguments \cite{klebanov2008} that in confining gauge theories with $N_c$ colors
the EE has a non-analyticity in the large $N_c$ limit: dependence of EE on $N_c$ suddenly jumps from $N_c^2$ to $N_c^0$ as 
the size of the subsystem crosses a critical value. This suggestion was confirmed numerically \cite{Buividovich2008} for a 
$\mathbb{Z}_2$ gauge theory, where the authors found an intrinsic ambiguity in the definition of the EE in the presence of 
gauge fields. It was shown that it is impossible to separate the Hilbert space into a tensor product of two Hilbert spaces 
without violating the Gauss law on the boundary of the bi-partition. In \cite{Buividovich2008} it was suggested to use a 
\textquotedblleft minimal way" of violating the Gauss law just in the border of the partition to compute the EE in a lattice 
gauge theory. This construction was formalized \cite{donelly2012} for the spin-network states and arbitrary gauge groups.

In the physics of black holes the EE of the matter fields in a gravitational background coincides with the entropy of 
the black hole for scalars and fermions \cite{Kabat1995}. In the same reference it was found that in the presence of
gauge fields there is an additional contact term that makes the EE negative for spatial dimension $D<8$ and thus cannot be 
interpreted as EE of any quantum field theory. The term was written as sum over trajectories starting and ending on the 
horizon. This observation carries a close resemblance with the construction of \cite{Buividovich2008}.

In condensed matter physics the (topological) EE can be used to classify the topological phases 
\cite{kitaev2006a,levin2006,isakov2011topological} of gapped systems. There is an intriguing possibility that EE can help 
to understand the gapless topological phases \cite{wenSRE}. 

The issue of ambiguity of the EE was recently addressed in \cite{Casini2013} where it was found that different extensions 
of the physical Hilbert space lead to different values of the (topological) EE. Nevertheless, it was concluded that the
ambiguity becomes irrelevant in the continuum limit and the usual replica method gives the correct answer.

In this letter we study the $U(N)$ Yang Mills theory in 1+1 dimensions (YM$_2$). This theory is exactly solvable and 
superrenormalizable. The simplicity of the model is explained by the large symmetry group of area-preserving diffeomorphisms 
that prevents the existence of any local degree of freedom. The model can be interpreted as a closed string theory \cite{Gross1993} 
and in the large $N$ limit the partition function can be written as a sum over the Riemann surfaces. It was used as a 
toy model to test the relationship between large-$N$ QCD and the (free) string theory. We take advantage of the simplicity 
of the model and derive expressions for the EE of YM$_2$ with $U(N)$ gauge group on a Riemann surface of genus $g$ for 
arbitrary bi-partition. We derive the Entanglement spectrum and the large $N$ limit for the EE.
\section{Definitions}
\subsection{Entanglement entropy of a partition}
The system under consideration is defined on one spatial and one (euclidean) time dimensions.
We will take the subsystem $A$ to be a union of $l$ disjoint intervals in the spatial dimension. In the text we will refer to it as $l$ cuts.
We denote the complement of $A$ by $\bar{A}$.

The Entanglement entropy (EE) is defined as the von Neumann entropy of the reduced density matrix $\rho_A={\rm tr}_{\bar{A}}\rho$ by,
\be
S = -{\mbox{tr}}\rho_A\ln\rho_A.
\ee
More generally, we define Renyi entropy for any integer $n$.
\be\la{ren}
S_n = \frac{1}{1-n}\ln{\mbox{tr}}\rho^n_A.
\ee
Then the EE is obtained from the Reyni entropy by
\be\la{trick}
S=-\lim_{n\rightarrow 1+0}\frac{\p}{\p n} {\mbox{tr}}\rho^n_A.
\ee
The subtle point in (\ref{trick}) is the definition of the derivative with respect $n$. In principle, if there are no 
divergences then every eigenvalue of $\rho_A$ is smaller than 1 and ${\mbox{tr}}\rho^n_A$ is absolutely convergent making the analytic continuation to any real $n$ easy. In reality, there are divergences and we have to deal with the analytic 
continuation to the real  $n$ on the case by case basis.

\subsection{Replica method}

A standard way to compute EE is the replica method \cite{calabrese2004a,holzhey1994a}. In order to compute ${\mbox{tr}}\rho^n_A$ one replaces the base manifold of the quantum field theory $\Sigma$ by 
its $n$-sheeted ramified covering $\Sigma_n$, with ramification points being the end points of the $l$ cuts. We denote the partition functions of the QFT on $\Sigma$ and on $\Sigma_n$ 
as $Z$ and $Z_n$ correspondingly. It was shown \cite{calabrese2004a} that 
\be\la{rep}
 {\mbox{tr}}\rho^n_A = \frac{Z_n}{Z^n}.
\ee
Thus the computation of the entanglement entropy is reduced to the computation of the partition function on $\Sigma_n$. EE is given by
\be\la{replica}
S=-\lim_{n\rightarrow1}\frac{\p}{\p n} \frac{Z_n}{Z^n} = \ln Z - \frac{1}{Z}\lim_{n\rightarrow1}\frac{\p}{\p_n}Z_n.
\ee

\subsection{Pure Yang-Mills theory in 2 dimensions}
Two dimensional Yang Mills (${\rm YM}_2$) is an exac\-tly solvable, super-renormalizable model and has been studied 
extensively \cite{Rusakov1990, Cordes1995}.  The model is fully specified by the choice of a gauge group $G$ and a Riemann surface $\Sigma$. The action is given by
\be\la{YM}
S[A] = \frac{1}{4e^2}\int_{\Sigma}d^2x\sqrt{g}g^{\mu\lambda}g^{\rho\nu}\tr F_{\mu \nu} F_{\rho \lambda},
\ee
where $F_{\mu \nu}^a = \partial_\mu A_\nu^a-\partial_\nu A_\mu^a+f^{abc}A_\mu^bA_\nu^c$ is the field strength tensor associated with the gauge group $G$ and $g_{\mu\nu}$ is the metric on the Riemann surface $\Sigma$. The action (\ref{YM}) is invariant with respect to area-preserving diffeomorphisms \cite{Cordes1995}. This large symmetry group is responsible for the simplicity of the theory. 

The partition function is given by
\be\la{pf}
Z = \int \mathcal{D}A e^{-S[A]}.
\ee
Notice that both the action and the partition function are invariant with respect to the transformation
\be
e\rightarrow\sqrt{r}e \quad g_{\mu\nu}\rightarrow rg_{\mu\nu}.
\ee
This implies that the coupling constant $e$ and the area $A$ will always enter together, so from now on we absorb $e$ into 
the definition of the area.
\subsection{Partition function}
Let us fix $G$ to be $U(N)$ or $SU(N)$. The partition function can be computed explicitly \cite{Rusakov1990,Migdal1975}
\be\la{action}
Z(A,g) = \sum_{R} (d_R)^{\chi(\Sigma)} e^{-\frac{A}{2N}C_2(R)},
\ee
where the sum runs over all irreducible representations of the gauge group, including the trivial one \cite{Rusakov1990}. Here, $C_2(R)$ is the 
quadratic Casimir, $d_R$ is the dimension of the representation $R$, $N$ is the t'Hooft coupling constant and 
$\chi(\Sigma)$ is the Euler characteristic of the Riemann surface $\Sigma$. 

In general, one encounters divergences when computing the partition function (\ref{pf})  in the presence of the external gravitational field. There are only two local counter terms needed to cancel them \cite{Witten1991}. 
\bea
I_1 &=& u\int d^2x \sqrt{g} \\
I_2 &=&-\frac{v}{2\pi}\int d^2x \sqrt{g}R
\eea
So the partition function has to be modified accordingly
\be\la{uvPF}
Z'(A,\Sigma,u,v) = e^{uA-v\chi(\Sigma)}Z(A,\Sigma).
\ee
Thus taking the regularization scheme into account gives a $2$-parametric family of theories with $u$ and $v$ being parameters.
\section{Entanglement entropy in ${\rm YM}_2$ theory}
We are going to merge the results reviewed in the previous section. Since the partition function depends only on the total area and the Euler characteristic of the Riemann surface. Thus application of the replica method is straightforward.

In order to compute $Z_n$ we need to know the Euler characteristic $\chi(\Sigma_n)$ of the $n$-sheeted ra\-mi\-fied covering $\Sigma_n$. This is given by Riemann-Hurwitz theorem. We have $2l$ ramification points of degree $n$.
\be\la{chin}
\chi(\Sigma_n) = n\chi(\Sigma) -2l(n-1) ,
\ee
Direct application of the replica (\ref{replica}) gives
\be\la{rhon}
\tr\rho^n_A = e^{-vl(2-2n)}\frac{\sum_R d_R^{n\chi(\Sigma) -2l(n-1)}e^{-An C_2(R)/2N}}{\left(\sum_R d^{\chi(\Sigma)}_R e^{-AC_2(R)/2N}\right)^n},
\ee

Already at this point we see dependence on $u$ canceled as it always happens in using the replica trick.
Combining (\ref{trick}) and (\ref{rhon}) we get the EE
\begin{multline}\la{answ}
S= 2lv + \ln\left(\sum_R d_R^{\chi(\Sigma)}e^{-\frac{A}{2N} C_2(R)}\right) - \\
-\frac{\sum_R d_R^{\chi(\Sigma)}e^{-\frac{A}{2N} C_2(R)}\ln(d_R^{\chi(\Sigma)-2l}e^{-\frac{A}{2N} C_2(R)})}{\sum_R d_R^{\chi(\Sigma)}e^{-\frac{A}{2N} C_2(R)}} .
\end{multline}
This formula is the first major result of this letter. An special case of this expression for $l=1$ and $v=0$ was obtained in \cite{velytsky2008}.

This can be written in compact form as follows. For any operator $X$, that is diagonal in the character basis we 
introduce the notation
\be
\langle X \rangle = \frac{\sum_R d_R^{\chi(\Sigma)}e^{-\frac{A}{2N} C_2(R)}X_R}{\sum_R d_R^{\chi(\Sigma)}e^{-\frac{A}{2N} C_2(R)}},
\ee
where $X_R = \langle R| X |R\rangle$ is the eigenvalue of $X$ on the state labeled by the irrep $R$. Then we can rewrite (\ref{answ}) as 
\be
S = 2lv + \ln Z - \langle \ln(d_R^{\chi(\Sigma)-2l}e^{-\frac{A}{2N} C_2(R)})\rangle.
\ee

\subsection{Torus}
In order to lighten up the notations and to give the area $A$ a thermal interpretation we choose $\Sigma$ to be a torus. 
This corresponds to a gauge theory with periodic boundary conditions in a thermal bath. In this case $\chi(\Sigma)=0$. We 
take one of the radii to be $\frac{1}{T}$ and the other one to be $1$.

Thus the entanglement entropy is given by
\begin{multline}\la{answT}
S = 2lv + \ln\left(\sum_R e^{-\frac{1}{2TN} C_2(R)}\right) - \\
-\frac{\sum_R e^{-\frac{1}{2TN} C_2(R)}\ln(d_R^{-2l}e^{-\frac{1}{2TN} C_2(R)})}{\sum_R e^{-\frac{1}{2TN} C_2(R)}}, 
\end{multline}
where we also replaced $A$ by $\frac{1}{T}$.

\subsection{Strong coupling}
The sum over irreps is essentially a strong coupling (low temperature) expansion. In the strong coupling limit 
we keep contributions from the trivial and the fundamental representations. Qua\-dra\-tic Casimir is normalized such that $C_2(\Box)=N$ and $d_{\Box}=N$. We have
\be
S(T) = 2lv + e^{-\frac{1}{2T}}(1+2l\ln N +\frac{1}{2T})+\ldots
\ee
At this point we make two observations.

 First, at zero temperature the entanglement entropy is completely determined by the regulator
\be
S(0) = 2lv,
\ee

Second, since the mass gap in the problem is set by the eigenvalues of the quadratic Casimir $m_{gap}=\frac{C_2(\Box)}{2N}$ the EE scales as
\be\la{ST}
S(T) - S(0) = O\left(\frac{m_{gap}}{T}e^{-\frac{m_{gap}}{T}}\right).
\ee
This scaling is similar to the scaling found for the matter fields \cite{Herzog2013,Herzog_boson,Cardy2014}, but  with additional factor of $\frac{1}{T}$. This factor is natural since the entropy is essentially an average of the logarithm of the density matrix and it goes as the Boltzmann weight times $\frac{1}{T}$. The eq. (\ref{ST}) is another new result of this paper.
\subsection{Weak coupling}
Weak coupling limit is more tricky. In this limit the sum over representations is not well behaved. Of course, we expect 
the entanglement entropy to be divergent at high temperature, as it should approach the scaling of the thermal entropy.
To observe this scaling, we take a simple gauge group. Let $G=SU(2)$. Representations of $SU(2)$ are labeled by an 
integer $m$. We have
\be
d_R= m,\quad C_2(R)= \frac{m^2-1}{2}.
\ee
The partition function is given by
\be
Z = \sum_{m=1}^{\infty} e^{-\frac{m^2-1}{8T}}=\frac{e^{\frac{1}{8T}}}{2}\left(\vartheta_3(0,e^{-\frac{1}{8T}})-1\right),
\ee
where $\vartheta_3$ is a Jacobi theta function. In the weak coupling limit ($T>>1$) EE scales as

\be
S(T)=\frac{3}{2}\ln{T}+\frac{1}{2}\ln{8\pi} -\gamma +O\left(\frac{\ln{T}}{T^{1/2}}\right),
\ee
where $\gamma$ the Euler-Mascheroni constant. 
\subsection{Higher genus}
The situation is a little different for genus $g>1$. First of all, the quantity (\ref{replica}) does not immediately have the EE meaning, but we will retain the same terminology.

 In this case the eq. (\ref{rhon}) still holds, with one important 
difference: $d_R$ enters the sum with a negative power, thus making the sum over $R$ absolutely convergent. The low 
temperature (large area) behavior is found in the same way as before.

In the zero-area limit YM$_2$ theory becomes a TQFT. The entanglement entropy is given by
\be\la{topS}
S^{top}_l = 2lv + \ln Z^{top} + \frac{2(l+g-1)}{Z^{top}}\sum_Rd_R^{2-2g}\ln d_R 
\ee
where we have denoted
\be
Z^{top} = \sum_R d_R^{2-2g}
\ee
Note, that sum in (\ref{topS}) is convergent, but this is not surprising as the area does not have a thermal interpretation 
anymore. 

\subsection{Entanglement Spectrum on the torus}
Using the previous results we can also study the eigenvalues of the logarithm of the partial density matrix, the so 
called entanglement spectrum \cite{Li2008}. It is introduced as follows. 
Given a reduced density matrix $\rho_A$ we can define an operator. 
\be\la{EH}
\rho_A= e^{-H_{e}}
\ee
 If we denote the eigenvalues of $\rho_A$ by $\Lambda_R$ and their degeneracies by $g_R$ then we have 
\begin{equation}
 {\rm tr}\rho_A^n=\sum_Rg_R\Lambda_R^n,
\end{equation}
comparing with (\ref{rhon}) we obtain ($\lambda=A/2N$)
\begin{eqnarray}
 \Lambda_R&=&\frac{\exp\left(-\lambda C_2(R)-2l\ln d_R\right)}{\sum_Re^{-\lambda C_2(R)}},\\
 g_R&=&d_R^{2l}
\end{eqnarray}
we find then that the eigenvalues of $H_e$ (the entanglement spectrum) are
\begin{equation}
\xi_R= \lambda C_2(R)+(2l-\chi(\Sigma))\ln d_R,
\end{equation}

\noindent each of them with degeneracy $g_R$.

\section{Large $N$ limit}
In two dimensions the large $N$ Yang-Mills theory on a sphere has the celebrated $3$-rd order Douglas-Kazakov phase transition \cite{Douglas1993}. For the other base manifolds the large $N$ limit is either not 
well behaved (for $\chi(\Sigma)<0$) or trivial (for $\chi(\Sigma)=0$). In this section we will study the EE in 
the large $N$ limit on a sphere.

\subsection{Free energy in large $N$}
In order to take the large $N$ limit we make group theory explicit. This procedure is discussed in great detail in 
\cite{Rusakov1990,Cordes1995,Douglas1993,Rusakov1993} so we will be very brief.

The irreps of $U(N)$ are labeled by non-decreasing integers $n_1\leq n_2\leq\ldots\leq n_N$. Quadratic 
Casimir and the dimensions of the irreps are parametrized as follows.
\bea
C_2(R) &=& \sum_{i=1}^Nn_i(n_i-2i+N+1),\\
d_R &=& \prod_{i<j}\left(1-\frac{n_i-n_j}{i-j}\right).
\eea
Then introducing $x = \frac{i}{N}$, $n(x) = \frac{n_i}{N}$, $h(x) = -n(x) + x - \frac{1}{2},$ and
\be
\rho(x) = \frac{dx}{dh},
\ee
\noindent we write the partition function on arbitrary Riemann surface of Euler characteristic $\chi$ as
\be\la{NZ}
Z(A,\chi) = e^{N^2F(A)} = \int D\rho e^{-N^2S^\chi_{eff}},
\ee
where
\begin{multline}
S^\chi_{eff} = \frac{A}{2}\int_0^1 dh \rho(h)h^2 \\
- \chi\int dh\int ds\rho(h)\rho(s) \ln|h-s| - \frac{A}{24} - \frac{3}{2}\chi.
\end{multline}

We are keeping $\chi$ arbitrary because we want to apply the expressions to the partition function on $\Sigma_n$. 
Integral in (\ref{NZ}) can be evaluated via the saddle point approximation. Since $n_i$'s are ordered $\rho(x)\leq 1$. This restriction is responsible for the DK phase transition.

\subsection{Sphere}
We take $\chi>0$ and solve the saddle point equation \cite{Douglas1993,Rusakov1993}.
\be\la{Saddle_point}
\rho(h) = \frac{A}{\pi} \frac{1}{\chi}\sqrt{\frac{2\chi}{A}-h^2}.
\ee
Evaluating the action on this solution we find
\be\la{Zext}
Z(A,\chi) = \exp N^2\left[\frac{\chi}{2}\ln\frac{\chi}{2A} + \chi + \frac{A}{24}\right].
\ee
Notice that the last two terms in this expression are precisely of the form (\ref{uvPF}).

We first compute the Renyi entropy $S_n$ for $l=1$. Combining (\ref{replica}),(\ref{chin}) and (\ref{Zext}) and setting $\chi(\Sigma)=2$ we have
\be
S_n  = N^2\left(\ln \frac{1}{A} + 1 - \frac{\ln n}{1-n}\right),
\ee
and the EE is given by
\be\la{largeNTS}
S= N^2\ln \frac{1}{A},
\ee
These expressions is another new result of this letter.

For general $l$ we encounter an issue. The Euler characteristic doesn't stay po\-si\-tive for arbitrary $n$, that is $\chi(\Sigma_n)>0$ implies $n<\frac{l}{l-1}$.
So if we proceed as before, our expression will be valid only for $n$ slightly bigger than $1$. This raises a 
question about the validity of the naive analytic continuation. Nonetheless, a sensible answer is obtained
\be\la{lS}
S=lN^2\ln \frac{1}{A}.
\ee
The end points of the cuts simply add additional de\-grees of freedom and the EE gets a contribution.

We point out that a similar issue arises on a torus in the large $N$ limit with the difference that for any $n>1$ the Euler 
characteristic changes sign and the analytic continuation is impossible. This leads to a divergence of the limit (\ref{trick}). Even though all Renyi entropies are well defined, the analytic continuation to EE is impossible.

The expressions (\ref{largeNTS}) and (\ref{lS}) are valid only for weak coupling (small area) since at some value of area $A_{cr}$ the condition $\rho(x)\leq1$ is violated.

\subsection{Douglas-Kazakov phase transition}

The EE is sensitive to the DK phase transition. After some gymnastics with elliptic functions (in line with \cite{Douglas1993})
we find that near the critical point $x_c=\pi^2$ we have for $\Delta S=S_{strong}(x)-S_{weak}(x)$ (with $x=A/\chi$)
 \be
 \frac{\Delta S(x)}{N^2}=-2l\left(\frac{x-x_c}{\pi^2}\right)^2\left[\frac{2x+x_c}{3\pi^2}\right] +\dots
 \ee
thus the entanglement entropy also has a critical point at $x=x_c$. In this point although the EE is continuous, but
its 2nd derivative is discontinuous. This is a signal of the DK phase transition.

\section{Conclusions}
We have computed the entanglement entropy of 2D pure Yang-Mills theory on a Riemann surface of arbitrary genus with any 
number of disjoint intervals in the bipartition. We found an exponential scaling at low temperatures which is similar to 
the ones found for the matter fields \cite{Herzog2013,Herzog_boson}. We have investigated the behavior of the EE near the 
DK phase transition and found that it has a critical point at the phase transition. For the higher genus Riemann surfaces 
the small area (topological) limit is convergent and the EE well defined. 

We did not find the non analytic behavior predicted in \cite{klebanov2008}, but it was expected as YM$_2$ does not know about any 
length scales except the size of the system. Another way to say it: all of the dependence of the EE on the length of the cut $L$ has the form of the Heaviside function $S\sim N^2\theta(L)$, so that $\frac{\partial S}{\partial L} \sim N^2 \delta(L)$, thus the critical 
value of $L$ is pushed to $0$ by the symmetries of the model.

It would be interesting to study the EE with additional matter fields and take advantage of the integrability of the 
Schwinger model. We leave this for the future work.

We thank A. Abanov, L. Fidkowski, C. Herzog, K. Jensen, F. Loshaj, N. Nekrasov for helpful discussions.

RS acknowledges the support of grant GIF 1167-165.14/2011.



\bibliographystyle{my-refs}
\bibliography{YangMills.bib}

\begin{thebibliography}{10}

\bibitem{amico2008a}
L.~Amico, R.~Fazio, A.~Osterloh, and V.~Vedral.
\newblock \emph{Entanglement in many-body systems}.
\newblock Rev. Mod. Phys., \textbf{80}, 517 (2008).

\bibitem{Callan1994}
C.~Callan and F.~Wilczek.
\newblock \emph{{On geometric entropy}}.
\newblock Physics Letters B, \textbf{333}, 55--61 (1994).

\bibitem{calabrese2004a}
P.~Calabrese and J.~Cardy.
\newblock \emph{Entanglement entropy and quantum field theory}.
\newblock J. Stat. Mech.: Theor. Exp., page P06002 (2004).

\bibitem{klebanov2008}
I.~R. Klebanov, D.~Kutasov, and A.~Murugan.
\newblock \emph{Entanglement as a probe of confinement}.
\newblock Nuclear Physics B, \textbf{796}, 274 -- 293 (2008).

\bibitem{Buividovich2008}
P.~Buividovich and M.~Polikarpov.
\newblock \emph{{Entanglement entropy in gauge theories and the holographic
  principle for electric strings}}.
\newblock Physics Letters B, \textbf{670}, 141--145 (2008).

\bibitem{donelly2012}
W.~Donnelly.
\newblock \emph{Decomposition of entanglement entropy in lattice gauge theory}.
\newblock Phys. Rev. D, \textbf{85}, 085004 (2012).

\bibitem{Kabat1995}
D.~Kabat.
\newblock \emph{{Black hole entropy and entropy of entanglement}}.
\newblock Nuclear Physics B, \textbf{453}, 281--299 (1995).

\bibitem{kitaev2006a}
A.~Kitaev and J.~Preskill.
\newblock \emph{Topological Entanglement Entropy}.
\newblock Phys. Rev. Lett., \textbf{96}, 110404 (2006).

\bibitem{levin2006}
M.~Levin and X.-G. Wen.
\newblock \emph{Detecting Topological Order in a Ground State Wave Function}.
\newblock Phys. Rev. Lett., \textbf{96}, 110405 (2006).

\bibitem{isakov2011topological}
S.~V. Isakov, M.~B. Hastings, and R.~G. Melko.
\newblock \emph{Topological entanglement entropy of a Bose-Hubbard spin
  liquid}.
\newblock Nature Physics, \textbf{7}, 772--775 (2011).

\bibitem{wenSRE}
X.~Chen, Z.-C. Gu, Z.-X. Liu, and X.-G. Wen.
\newblock \emph{Symmetry protected topological orders and the group cohomology
  of their symmetry group}.
\newblock Phys. Rev. B, \textbf{87}, 155114 (2013).

\bibitem{Casini2013}
M.~H. H.~Casini and J.~A. Rosabal.
\newblock \emph{Remarks on entanglement entropy for gauge fields}.
\newblock arXiv:1312.1183 [hep-th] (2013).

\bibitem{Gross1993}
D.~J. Gross and W.~T. IV.
\newblock \emph{Two-dimensional QCD is a string theory}.
\newblock Nuclear Physics B, \textbf{400}, 181 -- 208 (1993).

\bibitem{holzhey1994a}
C.~Holzhey, F.~Larsen, and F.~Wilczek.
\newblock \emph{Geometric and renormalized entropy in conformal field theory}.
\newblock Nucl.~Phys.~B, \textbf{424}, 443 (1994).

\bibitem{Rusakov1990}
B.~Rusakov.
\newblock \emph{Loop averages and partition functions in $U(N)$ Gauge theory on
  two dimensional manifolds}.
\newblock Mod. Phys. Lett A, \textbf{5}, 693--703 (1990).

\bibitem{Cordes1995}
S.~Cordes, G.~Moore, and S.~Ramgoolam.
\newblock \emph{Lectures on 2D yang-mills theory, equivariant cohomology and
  topological field theories}.
\newblock Nuclear Physics B - Proceedings Supplements, \textbf{41}, 184 -- 244
  (1995).

\bibitem{Migdal1975}
A.~Migdal.
\newblock \emph{Recursion equations in gauge field theories}.
\newblock Sov. Phys. JETP, \textbf{42(3)}, 413--418 (1975).

\bibitem{Witten1991}
E.~Witten.
\newblock \emph{{On quantum gauge theories in two dimensions}}.
\newblock Communications in Mathematical Physics, \textbf{141}, 153--209
  (1991).

\bibitem{velytsky2008}
A.~Velytsky.
\newblock \emph{Entanglement entropy in $d+1$ $SU(N)$ gauge theory}.
\newblock Phys. Rev. D, \textbf{77}, 085021 (2008).

\bibitem{Herzog2013}
C.~P. Herzog and T.~Nishioka.
\newblock \emph{Entanglement Entropy of a Massive Fermion on a Torus}.
\newblock arXiv:1301.0336 [hep-th] (2013).

\bibitem{Herzog_boson}
C.~P. Herzog and M.~Spillane.
\newblock \emph{Tracing through scalar entanglement}.
\newblock Phys. Rev. D, \textbf{87}, 025012 (2013).

\bibitem{Cardy2014}
J.~Cardy and C.~P. Herzog.
\newblock \emph{{Universal Thermal Corrections to Single Interval Entanglement
  Entropy for Conformal Field Theories}} (2014).

\bibitem{Li2008}
H.~Li and F.~D.~M. Haldane.
\newblock \emph{Entanglement Spectrum as a Generalization of Entanglement
  Entropy: Identification of Topological Order in Non-Abelian Fractional
  Quantum Hall Effect States}.
\newblock Phys. Rev. Lett., \textbf{101}, 010504 (2008).

\bibitem{Douglas1993}
M.~R. Douglas and V.~A. Kazakov.
\newblock \emph{Large N phase transition in continuum {\rm{$QCD_2$}}}.
\newblock Physics Letters B, \textbf{319}, 219 -- 230 (1993).

\bibitem{Rusakov1993}
B.~Rusakov.
\newblock \emph{{Large-N quantum gauge theories in two dimensions}}.
\newblock Physics Letters B, \textbf{303}, 95--98 (1993).

\end{thebibliography}







\end{document}